# Electromagnetic Pulse Scattering by a Spacecraft Nearing Light Speed

TIMOTHY J. GARNER,[1,*] AKHLESH LAKHTAKIA,[2] JAMES K. BREAKALL,[1] CRAIG F. BOHREN[3]

[1]*Department of Electrical Engineering, Pennsylvania State University, University Park, PA 16802*
[2]*Department of Engineering Science and Mechanics, Pennsylvania State University, University Park, PA 16802*
[3]*Department of Meteorology, Pennsylvania State University, University Park, PA 16802*
*\*Corresponding author: tjg236@psu.edu*

*Humans will launch spacecraft that travel at an appreciable fraction of the speed of light. Spacecraft traffic will be tracked by radar. Scattering of pulsed electromagnetic fields by an object in uniform translational motion at relativistic speed may be computed using the frame-hopping technique. Pulse scattering depends strongly on the velocity, shape, orientation, and composition of the object. The peak magnitude of the backscattered signal varies by many orders of magnitude depending on whether the object is advancing toward or receding away from the source of the interrogating signal. The peak magnitude of the backscattered signal goes to zero as the object recedes from the observer at a velocity very closely approaching light speed, rendering the object invisible to the observer. The energy scattered by an object in motion may increase or decrease relative to the energy scattered by the same object at rest. Both the magnitude and sign of the change depend on the velocity of the object, as well as on its shape, orientation, and composition. In some cases, the change in total scattered energy is greatest when the object is moving transversely to the propagation direction of the interrogating signal, even though the Doppler effect is strongest when the motion is parallel or antiparallel to the propagation direction.*

**OCIS codes:** *(290.5825) Scattering theory; (350.5720) Relativity; (280.1350) Remote sensing and sensors, backscattering*

## 1. INTRODUCTION

Humans will become a spacefaring species when technology allows [1]. Our unmanned probes and manned spacecraft will need to travel at relativistic velocities if they are to reach their destinations within short periods of time. In addition to these vehicles, planets, stars, and other astronomical objects such as relativistic jets [2] that travel at significant fractions of the speed of light will have to be radar tracked by space-traffic controllers. Accurate computation of electromagnetic scattering by the natural and human-made objects will be necessary for remote detection and tracking. Accurate computation of electromagnetic scattering would also be needed to design propulsion systems for laser-driven spacecraft [1]. Experience with our current spacecraft [3] strongly indicates that relativistic effects in electromagnetic scattering by objects moving at an appreciable fraction of the speed of light will not be negligible. Quantitative measures of scattering by a rapidly moving object may be more or less in magnitude than when it is at rest relative to the detector, and may change even when the object's velocity is transverse to the interrogating beam incident on the object.

"On the electrodynamics of moving bodies" is the title of Einstein's 1905 paper on what is now called the special theory of relativity; see Miller [4] and Lorentz *et al.* [5] for English translations of that famous paper. Einstein derived a law of reflection by a uniformly translating mirror, which is not quite the law known to Euclid around 300 BCE [6]. What is true for a mirror is true for any scatterer of electromagnetic signals.

More recently, scattering by other uniformly translating objects has been investigated using the frame-hopping technique. This requires four steps. First, the electric and magnetic fields of the interrogating (transmitted) signal are defined in an inertial frame $K'$ affixed to the transmitter. Second, the Lorentz transformation is used to determine the electric and magnetic fields of the interrogating signal in another inertial frame $K$ that is affixed to the object [7]. Third, the fields of the signal scattered in any specific direction in $K$ are computed using standard electromagnetic techniques [8-10]. Finally, the fields of the scattered signal are transformed back to $K'$, which is also the inertial frame followed by every detector of the scattered signal. The transmitter and the detector are taken to be co-located for monostatic radar.

Application of the Lorentz transformations in the second and fourth steps changes the spectrum, strength, propagation direction, and field orientations of the signal between the two inertial frames. Accordingly, the fields of the interrogating signal in $K$ depend on the object's velocity. The fields of the scattered signal in $K$ calculated for any direction in the third step depend on the interrogating signal in $K$ (and thereby the object's velocity) and on the object's shape, composition, and orientation. Finally, the fields of the scattered signals in $K'$ are altered from those in $K$. Details of the field transformations in the frame-hopping technique are given elsewhere [11].

Until very recently, computation of electromagnetic scattering by uniformly translating objects with relativistic velocities was limited to the scattering of monochromatic plane waves by objects with simple shapes, including half spaces [12], infinitely long cylinders [13,14], infinitely long wedges [15], and spheres [16-19]. Often but not always, the characteristic dimensions of the object would have to be small in comparison to the wavelength of the incident plane wave.

A monochromatic plane wave is, by definition, a sinusoidal function of time and exists for infinite duration. We recently implemented the frame-hopping technique to compute the more realistic time-limited signal backscattered by a uniformly translating sphere interrogated by a pulsed plane wave of finite duration [11]. We used both the analytical Lorenz–Mie theory [20] and the numerical finite-difference time-domain (FDTD) method to compute the scattered signal in $K$.

As objects can have complex shapes and material properties, we explore the effects of velocity, shape, orientation, and composition on electromagnetic pulse scattering by computing the fields and energies of signals scattered by silicon-carbide (SiC) disks and rods in uniform translational motion. Of simple shapes, disks and rods would be components of any spacecraft and could also represent space debris. We also investigated backscattered signals from a model spacecraft either advancing toward or receding away from a co-located transmitter-receiver pair. The complex shape of the model spacecraft necessitated the use of a numerical method for scattering in $K$, for which we chose a commercial implementation [21] of the FDTD method. Limitations on computational time and memory storage prevented us from pursuing large-scale computations necessary for full-scale spacecraft, but the essential features of time-domain scattering by uniformly translating objects were realistically captured by our calculations. With the advent of larger and faster computing hardware, our computational procedure can be scaled up.

We use primed variables in $K'$ and unprimed variables in $K$. Vectors are denoted with boldface and unit vectors are accented with carets. The angles $\theta$ and $\theta'$ are measured from the $z$ and $z'$ axes, respectively, and the angles $\phi$ and $\phi'$ from the $x$ and $x$ axes to the vector's projection onto the $xy$ and $x'y'$ planes.

## 2. Computational Methods

We used XFdtd [21], a commercial implementation of the finite-difference time-domain (FDTD) method [10,22], to compute the far-zone scattered fields in $K$. We used a 0.1-μm mesh size for all three orientations of the SiC rod and for the $z$-oriented SiC disk. The mesh size was chosen as 0.15 μm for the $x$-oriented SiC disk and the $y$-oriented SiC disks, the coarser mesh being necessary for these two cases because of the large size of the disk compared to the rod and the long duration of the scattered signals relative to those from the $z$-oriented disk. We modeled the relative permittivity of silicon carbide using the Lorentz oscillator with parameters available from Bohren and Huffman [20].

The interrogating signal in $K'$ is a pulsed plane wave whose electric field is given by

$$\mathcal{E}'_i(\mathbf{r}',t') = \hat{\mathbf{x}} \cos\left[\left(t' - \frac{z'}{c}\right) 2\pi f'\right] \exp\left[\frac{\left(t' - \frac{z'}{c}\right)^2}{2\sigma'^2}\right] \quad (1)$$

and whose magnetic field is given by

$$\mathcal{B}'_i(\mathbf{r}',t') = \frac{\hat{\mathbf{y}}}{c} \cos\left[\left(t' - \frac{z'}{c}\right) 2\pi f'\right] \exp\left[\frac{\left(t' - \frac{z'}{c}\right)^2}{2\sigma'^2}\right]. \quad (2)$$

We approximate the amplitude of the incident signal as constant over the region where the scattering takes place. Thus, we have omitted the inverse-distance dependence that would appear in the far-zone electric and magnetic fields radiated by a radar system, and we have normalized the peak value of the electric field to unity. However, let us note that an arbitrary carrier wave can be expressed as an angular spectrum of plane waves [23].

The energy density of the interrogating signal in units of energy per area is

$$U'_{inc} = \int_{-\infty}^{\infty} \left| \mathcal{E}'_i(\mathbf{r}',t') \times \frac{\mathcal{B}'_i(\mathbf{r}',t')}{\mu_0} \right| dt'$$
$$= \int_{-\infty}^{\infty} \frac{|\mathcal{E}'_i(\mathbf{r}',t')|^2}{\eta_0} dt', \quad (3)$$

where $\eta_0$ is the impedance of free space and $\mu_0$ is the permeability of free space.

The electric field of the scattered signal in each direction $\hat{\mathbf{q}}'_s$ in $K'$ is given by

$$\mathcal{E}'_s(\hat{\mathbf{q}}'_s;\mathbf{r}',t') = \hat{\boldsymbol{\theta}}' g(\hat{\mathbf{q}}'_s;t' - r'/c)/r' + \hat{\boldsymbol{\phi}}' h(\hat{\mathbf{q}}'_s;t' - r'/c)/r', \quad (4)$$

and the corresponding magnetic field by

$$\mathcal{B}'_s(\hat{\mathbf{q}}'_s;\mathbf{r}',t') = \hat{\boldsymbol{\phi}}' g(\hat{\mathbf{q}}'_s;t' - r'/c)/cr' - \hat{\boldsymbol{\theta}}' h(\hat{\mathbf{q}}'_s;t' - r'/c)/cr', \quad (5)$$

where $g$ and $h$ are functions [11] calculated using XFdtd. The unit vector $\hat{\mathbf{q}}'_s$ indicates the functions are different in general for each scattering direction [11]. The scattered energy density in units of energy per solid angle is

$$U'_{sca}(\hat{\mathbf{q}}'_s) = \int_{-\infty}^{\infty} \frac{r'^2 |\mathcal{E}'_s(\hat{\mathbf{q}}'_s;\mathbf{r}',t')|^2}{\eta_0} dt', \quad (6)$$

and the total scattered energy is

$$W'_{sca} = \int_0^\pi \int_0^{2\pi} U'_{sca}(\theta',\phi') \sin\theta' \, d\phi' d\theta', \quad (7)$$

where $\theta'$ and $\phi'$ are the angles of $\hat{\mathbf{q}}'_s$ in a spherical coordinate system in $K'$. The total scattered energy cross section is defined as $C'_{sca} = W'_{sca}/U'_{inc}$.

We calculated the far-zone fields of the scattered signal in $K'$ over a grid of 15 points in $\theta' \in [0,\pi]$ by 16 points in $\phi' \in [0,2\pi]$. We used the evaluation locations of 15-point Gauss–Kronrod quadrature [24,25] to choose values of $\theta'$, and we equally spaced the 16 points in $\phi'$ starting at $\phi' = 0$. As with the interrogating signal, the directions of propagation of the scattered signal in $K'$ and $K$ are different. We used the Lorentz transformation to convert each of the chosen scattering directions from $(\theta', \phi')$ in $K'$ to $(\theta, \phi)$ in $K$, and we added a time-domain far-zone sensor in each of the directions in XFdtd to compute the scattered electric field. Thereafter, we converted the electric field of the scattered signal to $K'$ using the Lorentz transformation, as detailed in Section 2D of Garner *et al.* [11].

We computed the scattered energy density $U'_{sca}$ (in units of energy per solid angle) in each direction by using rectangular integration over time of the instantaneous scattered power density (with units of power per solid angle) in $K'$. We computed the total scattered energy cross-section $C'_{sca}$ by numerically integrating $U'_{sca}$ over $\theta'$ using the 15-point Gaussian–Kronrod quadrature rules and over $\phi'$ using rectangular integration [24,25].

We used the difference between the 15-point Gauss–Kronrod and the 7-point Gauss–Legendre quadrature rules [24] to test for convergence of the integral of $U'_{sca}$ over $\theta'$. The greatest percent difference between the two rules was 2.8 % for the disk with $\hat{\mathbf{n}} = \hat{\mathbf{z}}$ when $\mathbf{v} = 0.2c\hat{\mathbf{z}}'$. The difference between the rules was less than 0.5 % for all other cases examined. Typical errors were on the order of $10^{-3}$ % for the rod and $10^{-2}$ % for the disk. We tested the convergence of the rectangular integration over $\phi'$ by comparing this result with rectangular integration over a subset of 8 equally spaced points in $\phi'$.

The greatest differences between the 8- and 16-point integration over $\phi'$ were 2.3 % and 2.5 % for the disk with $\hat{\mathbf{n}} = \hat{\mathbf{y}}$ for $\mathbf{v} = -0.2c\hat{\mathbf{z}}'$ and $\mathbf{v} = 0.2c\hat{\mathbf{y}}'$, respectively. The difference was less than 1.1 % for all other cases examined. Typical error estimates were on the order of $10^{-1}$% for both the rod and the disk.

The model spacecraft was taken to be made of a material with infinite electrical conductivity for calculations using XFdtd. The mesh size was 0.6 mm, except for (i) $\mathbf{v} = -0.9c\hat{\mathbf{z}}'$ when the mesh size was 0.24 mm and (ii) $\mathbf{v} = 0.99c\hat{\mathbf{z}}'$ when the mesh size was 1.5 mm. The finer mesh was necessary when $\mathbf{v} = -0.9c\hat{\mathbf{z}}'$ because the Doppler effect shifts the interrogating signal much higher in frequency in $K$ than in the other cases. XFdtd requires the mesh size to be one tenth of a wavelength or smaller. The coarser mesh was used when $\mathbf{v} = 0.99c\hat{\mathbf{z}}'$ because the duration of the backscattered signal was much greater than for the other cases. The coarser mesh allowed XFdtd to use a longer timestep, which reduced the computation time required relative to a shorter timestep.

## 3. RESULTS AND DISCUSSION

We begin with numerical results for pulsed-plane-wave scattering by individual circular rods and disks and then move on to pulse scattering by a spacecraft comprising rods, oblate spheroids, and parallelepipeds.

### A. Uniformly translating SiC rod and disk.

The orientation of a rod or a circular disk in the frame $K$ is specified by a unit vector $\hat{\mathbf{n}}$ along the symmetry axis, as shown in Fig. 1A for the rod and Fig. 1B for the disk. For calculations, the diameter and the thickness of the disk at rest are 20 µm and 1 µm, respectively, whereas the diameter and the length of the rod at rest are 1 µm and 10 µm, respectively. Fig. 1C shows the frame $K'$ along with its Cartesian coordinate axes as well as the angles $\theta'$ and $\phi'$; the direction of the electric field $\hat{\mathbf{E}}'_i$ and the direction of propagation $\hat{\mathbf{q}}'_i$ of the interrogating signal in $K'$; and the frame $K'$ along with its Cartesian coordinate axes. The vector $\mathbf{v}$ is the velocity of $K$ with respect to $K'$, and it is assumed that the origins and the coordinates axes of both frames coincide at time $t' = t = 0$. Figs. 1D, 1E, and 1F show the direction of the electric field $\hat{\mathbf{E}}_i$ and the direction of propagation $\hat{\mathbf{q}}_i$ in $K$ for $\mathbf{v} = \beta c\hat{\mathbf{z}}'$, $\mathbf{v} = 0.2c\hat{\mathbf{x}}'$, and $\mathbf{v} = 0.2c\hat{\mathbf{y}}'$, respectively, where $c$ is the free-space speed of light and the fraction $\beta \in (-1,1)$. When $\mathbf{v} = \beta c\hat{\mathbf{z}}'$, the propagation direction and electric field direction are unchanged between the two frames. However, when the velocity is transverse to the propagation direction of the interrogating signal in $K'$, the propagation directions in $K'$ and $K$ are different. As an example, $\hat{\mathbf{q}}_i$ is rotated by 11.5° with respect to $\hat{\mathbf{q}}'_i$ when $\beta = \pm 0.2$ and $\mathbf{v} \perp \hat{\mathbf{z}}'$. If $\mathbf{v}$ has a component in the $x'$ direction, the direction of the electric field vector will also be altered between the two frames, as shown in Fig. 1E.

The interrogating signal in $K'$ is a plane wave that is amplitude-modulated by a Gaussian function of time $t'$. It travels in the $+z'$ direction with its electric field pointing in the $x'$ direction, the frequency of the carrier plane wave is $f' = 20$ THz (carrier wavelength $\lambda' = 10$ µm), and the width parameter of the Gaussian function is $\sigma' = 0.05$ ps. Fig. 2A shows the interrogating signal in $K'$. We assumed that the pulsed plane wave is launched from a source located at $-D'\hat{\mathbf{z}}'$ in $K'$, where $D' >> c\tau'$ and $\tau' = 4\sigma'$ is the pulse duration. We also assumed that the object is located in the neighborhood of the origin of $K'$ at $t' = 0$, so that an interrogating beam of finite cross section could be approximated by a plane wave of infinite cross section for scattering calculations involving an object of finite dimensions. The implementation of the frame-hopping technique is described in detail elsewhere [11].

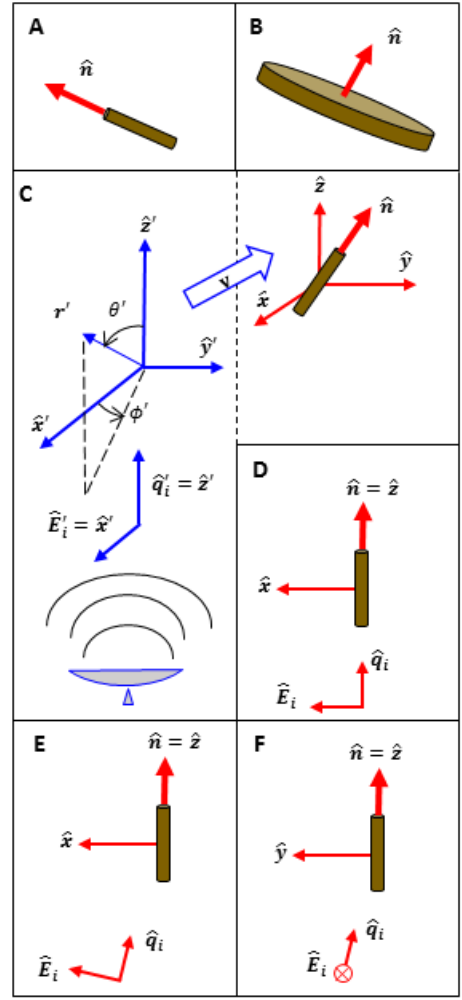

Fig. 1: Geometry of pulse-scattering problem. Vectors in $K'$ are blue, and those in $K$ are red. **(A)** SiC rod with its orientation vector $\hat{\mathbf{n}}$. **(B)** SiC disk with its orientation vector $\hat{\mathbf{n}}$. **(C)** The inertial frames K' and K, with K moving away from $K'$ with velocity $\mathbf{v}$, along with the directions of the electric field and the propagation of the interrogating signal denoted by $\hat{\mathbf{E}}'_i$ and $\hat{\mathbf{q}}'_i$, respectively, in $K'$. **(D)** The directions of the electric field and the propagation of the interrogating signal denoted by $\hat{\mathbf{E}}_i$ and $\hat{\mathbf{q}}_i$, respectively, in K when $\mathbf{v} = \beta c\hat{\mathbf{z}}'$. **(E)** Same as 1D but for $\mathbf{v} = 0.2c\hat{\mathbf{x}}'$. **(F)** Same as 1D but for $\mathbf{v} = 0.2c\hat{\mathbf{y}}'$. The electric field vector is pointing into the page.

Fig. 2 presents the time traces of the backscattered signals in $K'$ from the chosen SiC disk oriented such that $\hat{\mathbf{n}} = \hat{\mathbf{z}}$. For the backscattered signal, we took only the portion of the backscattered electric field co-polarized with the interrogating signal. The backscattered signal from the stationary disk is shown in Fig. 2B for reference. Figs. 2C and 2D display the backscattered signals when the disk is directly receding from and advancing toward the transmitter, respectively, with $\mathbf{v} = 0.2c\hat{\mathbf{z}}'$ and $\mathbf{v} = -0.2c\hat{\mathbf{z}}'$. Appreciable Doppler shifts relative to the backscattered signal from the stationary disk on the order of 50% are evident in both Figs. 2C and 2D. Figs. 2E and 2F show the backscattered signals when $\mathbf{v} = 0.2c\hat{\mathbf{x}}'$ and $\mathbf{v} = 0.2c\hat{\mathbf{y}}'$, respectively. Both of these backscattered signals appear very similar to, but not the same as, the backscattered signal from the stationary disk in Fig. 2B.

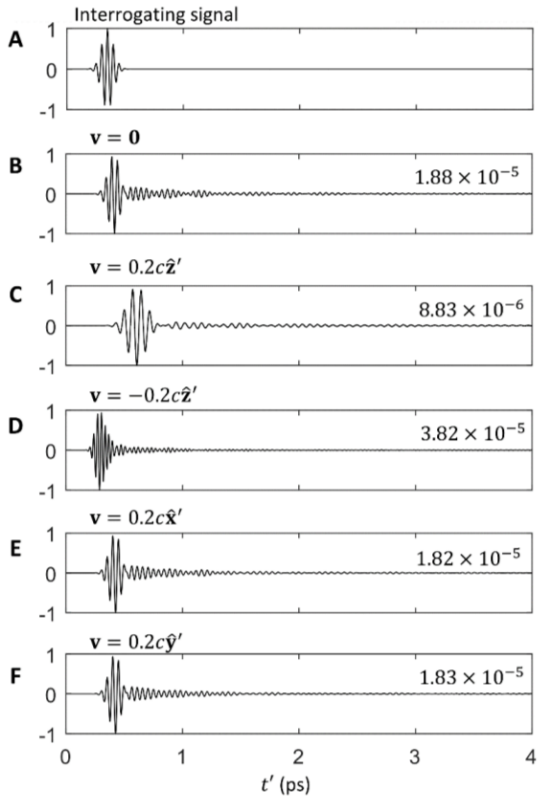

Fig. 2: Time traces of the interrogating and backscattered signals in $K'$ from the z-oriented SiC disk. (A) Interrogating signal. (B) Normalized backscattered signal from the stationary SiC disk. Signal is the $-x'$-directed component of the far-zone scattered electric field multiplied by the radial distance $r'$ from the origin and normalized such that its peak value is unity. Normalization constant is printed on the panel. (C) Same as 2B but for $v = 0.2c\hat{z}'$. (D) Same as 2B but for $v = -0.2c\hat{z}'$. (E) Same as 2B but for $v = 0.2c\hat{x}'$. (F) Same as 2B but for $v = 0.2c\hat{y}'$.

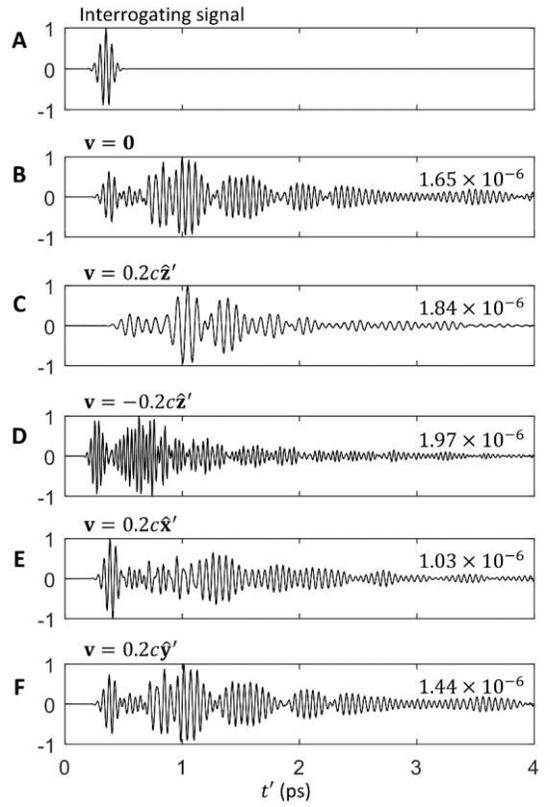

Fig. 3: Same as Fig. 2 but for the y-oriented SiC disk.

Fig. 3 shows the time traces of the backscattered signals from the SiC disk oriented such that $\hat{n} = \hat{y}$. The peak magnitudes of these signals are roughly an order of magnitude less than those from the z-oriented disk. The time traces of the backscattered signals from the y-oriented disk are much more complex and of greater duration than those from the z-oriented disk and strongly depend on the velocity. In $K$, the electromagnetic energy of the target is increased rapidly from zero by the interrogating pulse, but over a time much larger than its width the target continues to radiate this energy as a series of broader pulses with decreasing magnitude. For the stationary disk, the scattered signal's duration is approximately 4.0 ps, twenty times longer than the 0.2-ps duration of the incident pulse. When $v = 0.2c\hat{z}'$, the duration of the incident pulse in $K$ is increased to 0.244 ps by time dilation, and the duration of the scattered signal in $K$ is approximately 2.45 ps. Appreciable Doppler shifts relative to the stationary case are visible with $v = 0.2c\hat{z}'$ and $v = -0.2c\hat{z}'$, just as for the z-oriented disk in Fig. 2.

Fig. 4 shows the time traces of the backscattered signals for the SiC rod with $\hat{n} = \hat{x}$. As with the disk, Doppler shifts are readily apparent when $v = 0.2c\hat{z}'$ and $v = -0.2c\hat{z}'$. The duration of the time trace for $v = 0.2c\hat{z}'$ in Fig. 4C is less than the duration for other velocities because the FDTD simulation reached steady state and terminated more quickly than the simulations for the other velocities.

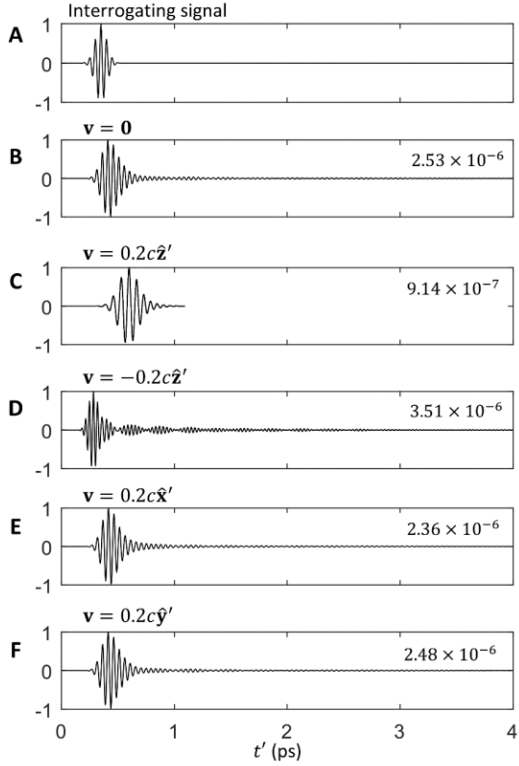

Fig. 4: Same as Fig. 2 but for the *x*-oriented SiC rod.

As the pulsed plane wave launched by the transmitter is of finite duration, it has a finite energy per unit area $U'_{inc}$. The scattered signal in any direction will have a finite duration. Hence, the total energy $W'_{sca}$ scattered in all directions must be finite. We therefore defined $C'_{sca} = W'_{sca}/U'_{inc}$ as the total scattered energy cross section with units of area. The procedure to calculate $W'_{sca}$ is presented in Section 3.

Values of $C'_{sca}$ for the SiC rod and disk are provided in Table 1 for $\hat{\mathbf{n}} \in \{\hat{\mathbf{x}}, \hat{\mathbf{y}}, \hat{\mathbf{z}}\}$ and $\mathbf{v} \in \{\mathbf{0}, 0.2c\hat{\mathbf{z}}', -0.2c\hat{\mathbf{z}}', 0.2c\hat{\mathbf{x}}', 0.2\hat{\mathbf{y}}'\}$. For all velocities considered, the rod has the largest $C'_{sca}$ when it is oriented along the *x* axis rather than along the *y* or the *z* axes. Thus, we conclude that $C'_{sca}$ is maximized if $\hat{\mathbf{n}}$ and $\hat{\mathbf{E}}_i$ are aligned parallel to each other. Likewise, in Table 1, the disk has the largest value of $C'_{sca}$ when it is oriented along the *z* axis rather than along the *x* or the *y* axes, so that $C'_{sca}$ is maximized if $\hat{\mathbf{n}}$ and $\hat{\mathbf{q}}_i$ are aligned parallel to each other.

The change in $C'_{sca}$ because of relative uniform motion usually is greatest when the object moves directly away from or towards the source of the interrogating signal ($\mathbf{v} = \pm 0.2c\hat{\mathbf{z}}'$). Velocities in these directions result in the largest changes in the magnitude and the spectrum of the interrogating signal between $K'$ and $K$ [11].

For the *z*-oriented SiC rod, the greatest increase in $C'_{sca}$ occurs when the velocity is in the *x* direction. This may be understood by referring to Fig. 1E, which shows that the electric field of the interrogating signal in $K$ has a component aligned with the rod's axis when the velocity is in the *x* direction. That alignment enhances scattering by the rod in $K$ [25]. For the other four velocities considered in Table 1, the electric field of the interrogating signal in $K$ is pointed in the *x* direction, normal to the rod's axis. We also computed the total energy scattered by the z-oriented rod moving with $\mathbf{v} = 0.5c\hat{\mathbf{x}}'$ to see if that would further increase $C'_{sca}$. For that velocity, $C'_{sca}$ is 34.9 times that of the stationary rod and 9.0 times that of the rod with $\mathbf{v} = 0.2c\hat{\mathbf{x}}'$.

For a given velocity, both the magnitude and the sign of the change in $C'_{sca}$ depend on the shape and orientation of the object in Table 1. For both the rod and disk oriented in the *x* direction, $C'_{sca}$ decreases when the object either advances towards or recedes from the transmitter, but $C'_{sca}$ increases when the object is moving in the *x* or *y* directions. For the *y*-directed rod, $C'_{sca}$ decreases when the rod recedes from the transmitter and increases when it advances towards it. However, the *y*-directed disk has a larger value of $C'_{sca}$ when it moves directly away from the transmitter and a smaller value of $C'_{sca}$ when it moves directly towards the transmitter. The value of $C'_{sca}$ increased relative to the stationary case for the *x*-, *y*-, and *z*-oriented rods with a velocity transverse to the propagation direction of the interrogating signal.

**Table 1: Total scattered energy cross sections for SiC disk and rod.** *A* is the physical cross section of the object at rest projected onto the *xy* plane. %Δ is the difference in $C'_{sca}$ from the stationary case for the given object and orientation.

| | | SiC Disk | | | SiC Rod | | |
|---|---|---|---|---|---|---|---|
| $\hat{\mathbf{n}}$ | $\mathbf{v}$ | $C'_{sca}$ (μm)² | $\frac{C'_{sca}}{A}$ | %Δ | $C'_{sca}$ (μm)² | $\frac{C'_{sca}}{A}$ | %Δ |
| $\hat{\mathbf{x}}$ | 0 | 139 | 6.95 | - | 67.1 | 6.71 | - |
| $\hat{\mathbf{x}}$ | $0.2c\hat{\mathbf{z}}'$ | 35.3 | 1.77 | −74.5% | 18.8 | 1.88 | −72.0% |
| $\hat{\mathbf{x}}$ | $-0.2c\hat{\mathbf{z}}'$ | 107 | 5.35 | −23.0% | 50.8 | 5.08 | −24.3% |
| $\hat{\mathbf{x}}$ | $0.2c\hat{\mathbf{x}}'$ | 166 | 8.30 | +19.4% | 67.2 | 6.72 | +0.1% |
| $\hat{\mathbf{x}}$ | $0.2c\hat{\mathbf{y}}'$ | 144 | 7.20 | +3.6% | 71.6 | 7.16 | +6.7% |
| $\hat{\mathbf{y}}$ | 0 | 324 | 16.2 | - | 0.401 | 0.04 | - |
| $\hat{\mathbf{y}}$ | $0.2c\hat{\mathbf{z}}'$ | 360 | 18.0 | +11.1% | 0.122 | 0.01 | −69.6% |
| $\hat{\mathbf{y}}$ | $-0.2c\hat{\mathbf{z}}'$ | 319 | 16.0 | −1.2% | 2.08 | 0.21 | +418.7% |
| $\hat{\mathbf{y}}$ | $0.2c\hat{\mathbf{x}}'$ | 324 | 16.2 | 0% | 0.475 | 0.05 | +18.5% |
| $\hat{\mathbf{y}}$ | $0.2c\hat{\mathbf{y}}'$ | 343 | 17.2 | +5.9% | 0.472 | 0.05 | +17.5% |
| $\hat{\mathbf{z}}$ | 0 | 641 | 2.04 | - | 0.590 | 0.75 | - |
| $\hat{\mathbf{z}}$ | $0.2c\hat{\mathbf{z}}'$ | 459 | 1.46 | −28.4% | 0.104 | 0.13 | −82.4% |
| $\hat{\mathbf{z}}$ | $-0.2c\hat{\mathbf{z}}'$ | 767 | 2.44 | +19.7% | 1.88 | 2.39 | +218.6% |
| $\hat{\mathbf{z}}$ | $0.2c\hat{\mathbf{x}}'$ | 650 | 2.07 | +1.4% | 2.28 | 2.90 | +286.4% |
| $\hat{\mathbf{z}}$ | $0.2c\hat{\mathbf{y}}'$ | 648 | 2.06 | +1.1% | 0.708 | 0.90 | +20.0% |

Based on the results for the *z*-oriented rod, we considered a rod with its axis aligned with the $\{\theta = 11.5°, \phi = 180°\}$ direction. For this orientation, the rod's axis is aligned with the direction of propagation of the interrogating signal in $K$ when $\mathbf{v} = 0.2c\hat{\mathbf{x}}'$. We hypothesized that $C'_{sca}$ would decrease relative to the stationary case when $\mathbf{v} = 0.2c\hat{\mathbf{x}}'$ and increase when $\mathbf{v} = -0.2c\hat{\mathbf{x}}'$. Indeed, we found that $C'_{sca}$ decreased by 65 % when $\mathbf{v} = 0.2c\hat{\mathbf{x}}'$ but increased by 277 % when $\mathbf{v} = -0.2c\hat{\mathbf{x}}'$ with respect to the value of $C'_{sca}$ for the stationary rod.

**B. Uniformly Translating Model Spacecraft**

We also did calculations for an 8.4-cm-long, 5.0-cm-wide metallic model spacecraft somewhat resembling USS Enterprise in the *Star Trek* television series [27]. Fig. 5A shows the model spacecraft advancing directly toward the transmitter. The interrogating signal in $K'$, shown in Fig. 5B, is a plane wave amplitude-modulated by a Gaussian function of time $t'$. This signal travels in the $+z'$ direction with its electric field pointing in the $x'$ direction, the frequency of the carrier plane wave is $f' = 20$ GHz, and the width parameter of the Gaussian function is $\sigma' = 0.05$ ns. The decrease in the carrier frequency from 20 THz (for the SiC rod and disk) allowed the spacecraft dimensions to considerably exceed those of the SiC rod and disk when using the chosen commercial implementation [21] of the FDTD method. Simultaneously, we could approximate the metal as a having infinite electrical conductivity, which is appropriate at frequencies well below the damping frequency of common metals [28]. Furthermore, the infinite-conductivity approximation allows the model spacecraft to serve as a scale model [29].

Figs. 5C, 5D, and 5E show the backscattered signals from the spacecraft facing the transmitter while stationary, advancing at $\mathbf{v} =$

$-0.2c\hat{z}'$, and advancing at $\mathbf{v} = -0.9c\hat{z}'$, respectively. The time scale varies from panel to panel. The peak magnitude of the backscattered signal increases with increasing magnitude of the velocity and is roughly three orders of magnitude greater for $\mathbf{v} = -0.9c\hat{z}'$ than for $\mathbf{v} = 0$. This tremendous difference shows that radar backscattering has the potential to measure spacecraft speed even of relativistic magnitudes. A monostatic radar has its transmitter and receiver co-located. Monostatic radar today determines automobile speed through the Doppler shift, i.e., the change in the frequency of the backscattering continuous-wave (not pulsed) signal from that of the incident continuous-wave signal [30]. Pulse radar uses arrival time to determine the range and the two-way Doppler shift to determine the radial velocity of the scattering object [31]. These objects move at very low speeds ($|\beta| \ll 1$). At relativistic speeds, the peak magnitude of the backscattered signal will provide additional information about the object.

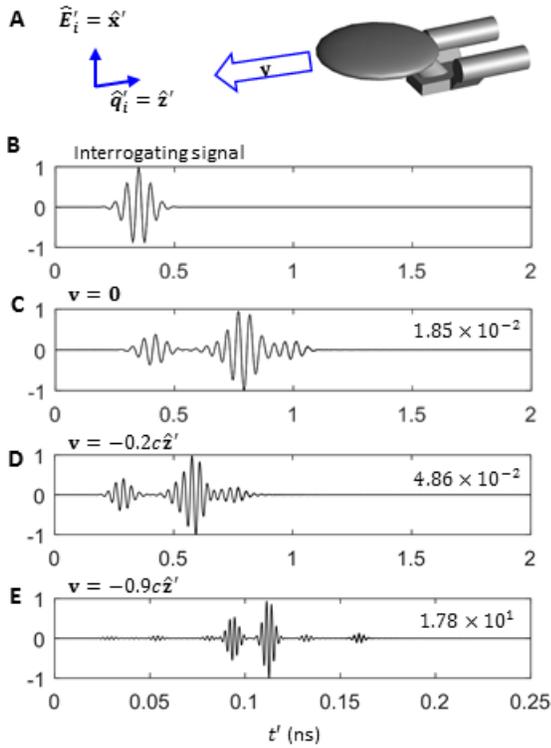

Fig. 5: Time traces of interrogating and backscattered signals in K' from model spacecraft. Time scales vary from panel to panel. (A) Interrogating signal. (B) Spacecraft model advancing toward the transmitter with velocity $\mathbf{v}$, along with propagation direction of interrogating signal $\hat{\mathbf{q}}_i'$ and direction of the interrogating electric field vector $\hat{\mathbf{E}}_i'$ in K'. (C) Normalized backscattered signal from stationary spacecraft. Signal is the $-x'$ component of the far-zone electric field multiplied by the radial distance r' from the origin and normalized such that its peak magnitude is unity. Normalization constant is printed on the panel. (D) Same as 4C but with spacecraft advancing toward source at $\mathbf{v} = -0.2\,c\hat{z}'$. (E) Same as 4D but with $\mathbf{v} = -0.9\,c\hat{z}'$.

Fig. 6 shows backscattered signals from a receding spacecraft. Fig. 6A shows the shows the model spacecraft directly receding from the transmitter, while Fig. 6B interrogating signal. Figs. 6C, 6D, 6E, and 6F show the backscattered signals from the spacecraft facing away from the transmitter while stationary, receding at $\mathbf{v} = 0.2c\hat{z}'$, receding at $\mathbf{v} = 0.9c\hat{z}'$, and receding at $\mathbf{v} = 0.99c\hat{z}'$, respectively. When the spacecraft recedes from the transmitter, the peak magnitude of the backscattered signal decreases with increasing speed. The peak magnitude when $\mathbf{v} = 0.9c\hat{z}'$ is about three orders of magnitude less than when the spacecraft is stationary, and it is five orders of magnitude less when $\mathbf{v} = 0.99c\hat{z}'$. Thus, as monostatic radar detection of a receding spacecraft would be much more difficult than that of an advancing spacecraft at the same uniform speed, a strategy to avoid detection by a monostatic radar would be to recede with a uniform speed close to the speed of light. Finally, even when the spacecraft is stationary, the shape of the backscattered signal depends on whether it is facing toward or away from the transmitter, as may be seen by comparing Figs. 5C and 6C.

When the uniformly translating object advances toward the source at a speed approaching $c$, the Doppler shift of the interrogating signal from K' to K grows unboundedly. This increases the electrical size of the object in K which demands more computational resources to compute the backscattered signal in K. For this reason, we were not able to compute the backscattered signal from the model spacecraft advancing toward the transmitter at $\mathbf{v} = -0.99c\hat{z}'$.

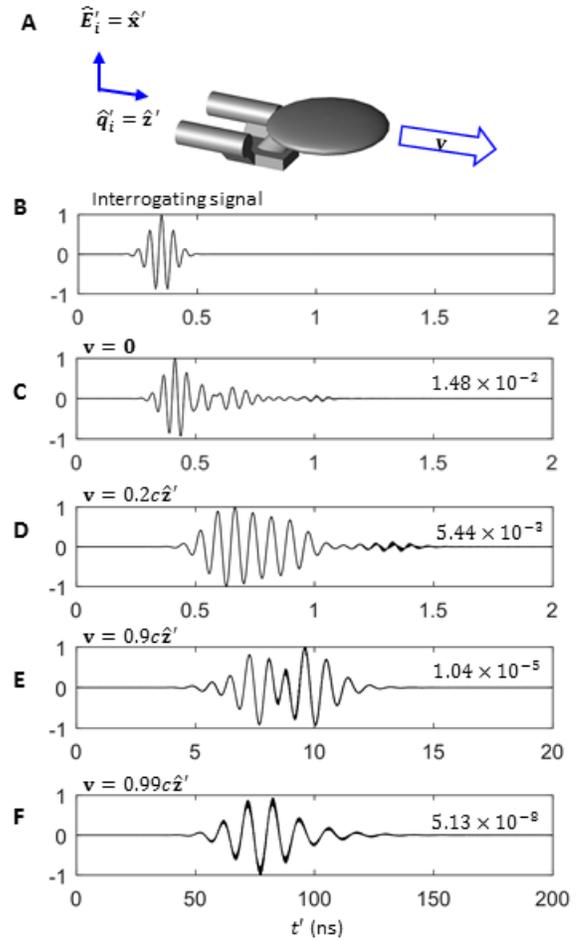

Fig. 6: Same as Fig. 5 but with spacecraft receding at (C) $\mathbf{v} = 0$, (D) $\mathbf{v} = 0.2c\hat{z}'$, (E) $\mathbf{v} = 0.9c\hat{z}'$, and (F) $\mathbf{v} = 0.99c\hat{z}'$

Fig. 7 provides time traces of the interrogating and the backscattered signals for two different combinations of the orientation of the model spacecraft and the direction of its velocity orthogonal to the direction of propagation of the interrogating signal, with the speed being fixed at $0.9c$. A close examination of these figures reveals that specific combinations of the orientation and the direction of motion can seriously distort the backscattered signal of a uniformly moving spacecraft compared to that of a stationary counterpart. A catalog of the specific features of the backscattered signal could be used to determine the make and the model of a spacecraft, its orientation, and its speed, similar to radar signature databases used for aircraft identification [32].

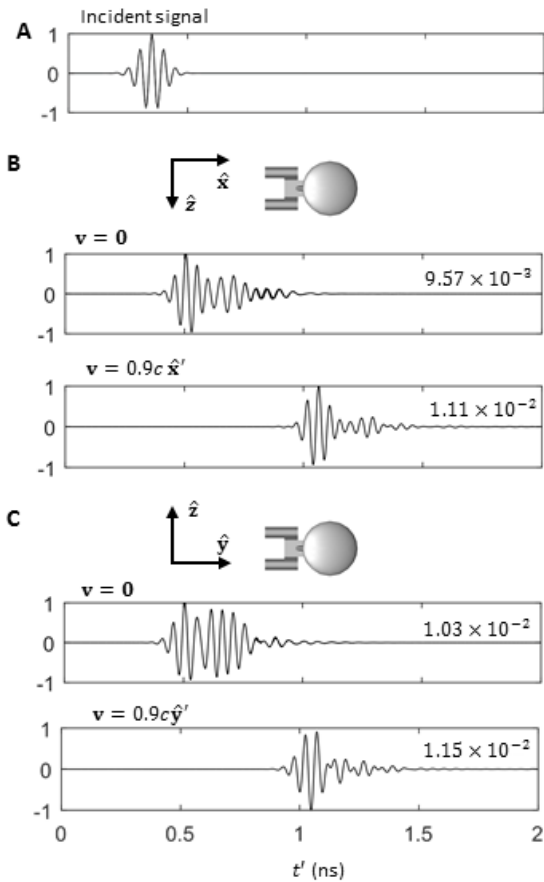

Fig. 7: Time traces of the interrogating and backscattered signals in K' from model spacecraft with various orientations. (A) Interrogating signal. (B)-(C) Normalized backscattered signal from the model spacecraft while stationary and with |**v**| = 0.9c. The velocity is printed above each time trace, and the orientation of the spacecraft is shown in each panel. Signal is the −x'-directed component of the far-zone scattered electric field multiplied by the radial distance r' from the origin and normalized such that its peak magnitude is unity. Normalization constant is printed on the panel.

## 4. CONCLUDING REMARKS

Both the amplitude and the shape of the backscattered signal depend on the orientation and velocity of the uniformly translating object, as is demonstrated by the backscattered signals from the SiC rod, SiC disk, and the metallic model spacecraft. For instance, when the spacecraft advances directly toward the transmitter at 0.9c, the backscattered signal is a series of pulses (Fig. 5E), each similar in shape to the interrogating signal but with varying amplitudes. When the spacecraft recedes directly from the transmitter at 0.9c, the backscattered pulses are much closer in time (Fig. 6E) and coalesce into a single pulse when recession occurs with 0.99c (Fig. 6F). The peak magnitude of the backscattered signal from a receding or advancing spacecraft decreases or increases rapidly as it approaches the speed of light, thereby providing a strategy to evade detection. Although the model spacecraft considered in our calculations is physically small, the same effects would be seen for a larger spacecraft as long as the frequency of the interrogating signal in K' is adjusted such that the size of the spacecraft in relation to the wavelength of the carrier plane wave is constant. There is no simple relationship between an object's velocity and its total scattered energy cross section $C'_{sca}$ relative to the object at rest. For a given velocity, $C'_{sca}$ may increase or decrease depending on its orientation. Moreover, $C'_{sca}$ can decrease or increase when the motion is transverse to the propagation direction of the interrogating signal. If the transformation from K' to K steers the interrogating signal into a null in the object's scattering pattern in K, $C'_{sca}$ can increase. This is demonstrated numerically by the SiC rod pointing in the {$\theta$ = 11.5°, $\phi$ = 180°} direction, whose cross section decreases when **v** = 0.2c$\hat{x}'$ compared to when the rod is stationary. In contrast, $C'_{sca}$ of the z-directed SiC rod with **v** = 0.2c$\hat{x}'$ increases relative to the stationary case.

The frame-hopping technique applies only to uniformly moving objects. Different techniques must be used to account for acceleration. Our analysis considered only scattering of a single pulse, but actual pulsed-radar systems typically transmit pulses periodically at a given repetition frequency [31]. Accommodation of multiple pulses in our computational approach is straightforward. The evolution of the backscattered signals from consecutive incident pulses will give additional information about the object's trajectory, including its acceleration. The accuracy of the scattered-energy computations is limited by the accuracy of the method used to calculate the scattered fields in K in the third step of the frame-hopping technique. With further advances in computer speed and memory availability, these limitations are expected to diminish.

## Acknowledgments

The authors thank Tarun Chawla and Walter Janusz from Remcom, Inc. (State College, PA, USA) for providing the XFdtd software and technical assistance with that software.

T. J. G. was supported by a Graduate Excellence Fellowship from the Pennsylvania State University College of Engineering. A. L. is grateful for the support of the Charles Godfrey Binder Endowment at the Pennsylvania State University.

This research was conducted with Advanced CyberInfrastructure computational resources provided by the Institute for CyberScience at The Pennsylvania State University (http://ics.psu.edu)